\documentclass[prc,aps,nofootinbib,showkeys,showpacs,twocolumn]{revtex4-2}   

\usepackage[T1]{fontenc} %Helpful for font encoding the output (useful for accents and copy/pasting text as a PDF)
\usepackage{epsfig}  
\usepackage{graphicx}  
\usepackage{amssymb}
\usepackage{amsmath}
\usepackage[english]{babel}
\usepackage{color}  
\definecolor{darkgreen}{rgb}{0.2,0.5, 0.2}

\usepackage{dcolumn,booktabs}
\newcolumntype{d}[1]{D{.}{.}{#1}}
\usepackage{lipsum}

\usepackage{dsfont}

\begin{document}

%\title{Microscopic description of fission fragment spin formation during and after scission}
\title{Microscopic description of the torque acting on fission fragments}
%\title{Microscopic description of fission fragment spin formation during and after scission}
%\title{Microscopic description of fission fragment spin formation during and after scission}

\author{Guillaume Scamps$^{1}$}
\email{gscamps@uw.edu} 
\address{$^{1}$ Department of Physics, University of Washington, Seattle, Washington 98195-1560, USA
} 

\begin{abstract} 
When two fragments are created in a fission decay, any torque due to nuclear and Coulomb interaction can change the fragment's angular momentum. This article explores the character and magnitude of the angular momentum as a function of the initial conditions around the scission point using the time-dependent Hartree-Fock theory.  To understand the torque acting on the fragments, the Frozen Hartree-Fock method is also used to determine the collective potential at scission. Two $^{240}$Pu fission channel  ( $^{132}$Sn+$^{108}$Ru and  $^{144}$Ba+$^{96}$Sr ) are studied. These two channels cover different shapes (spherical, quadrupole, and octupole deformation) of the fragments.
It is found that the angular momentum generated by the Coulomb interaction after fission is mainly collective, while this is not the case for the angular momentum generated at scission.
The competition between rotational modes (bending, wriggling, and twisting) is discussed and shows that the angular momentum is generated mainly perpendicular to the fission axis.
%twisting mode is less populated than the bending and wriggling mode.
\end{abstract}

\maketitle

%%%%%%%%%%%%%%%%%%%%%%%%%%%%%%%%%%%%%%%%%%%%%%%%%%%%%%%%%%%%%%%%%%%%%%%%%%%%%%%%
 
 \section*{Introduction}
 
 The generation of angular momentum in fission has several mechanisms at different stages of the fission process. One source is the fluctuations that build up during the shape evolution between the initial configuration and the scission \cite{Mor80,Dos85,Ran21,Ran22}. At the scission point, the fragments are deformed and tend to be aligned with the fission axis. Such polarization of the fragment creates an angular momentum  \cite{Mik99,Bon07} due to the uncertainty principle \cite{Fra04}. Once the fragments are separated, the Coulomb force can create a torque leading to an additional angular momentum \cite{Hof64,Ald75,Ber19}. This last effect is found to have large theoretical uncertainties  \cite{Mis99,Ber19,Ras69}.% Also, it is claimed that the experimental data \cite{Wil21} do not support the presence of this effect due to the absence of $Z_1 Z_2$ dependency. 
 
Recently, a new method has been proposed to extract the angular distribution microscopically in static \cite{Ber19b,Mar21} and dynamic \cite{Bul21,Bul22} density functional theory. While the projection is a powerful tool to extract exactly the distribution of a discrete observable in a quantum N-body wave function, the calculation still suffers from limitations due to the theory producing that state. In a mean-field dynamic calculation, the fluctuations of the mean-field are not taken into account which prevents the self-consistent description of the Coulomb-induced rotation. To overcome that limitation would require the use of a beyond mean-field method such as a stochastic treatment \cite{Bul19,Tan18} of the fission.

In this work, we assume that fluctuations build during the descent of the system from the saddle to the scission leads to a pair of fragments at a given deformation, distance, and orientation at scission. 
%All the sources of excitation energy (velocity and angular momentum) of the fragments are neglected.
The goal of the present calculation is to investigate the last stage of fission using the Frozen-Hartree-Fock (FHF) method \cite{Sim17,Uma21} (equivalent to the sudden approximation \cite{Den02,Was08,Sim08,Sim12}) and the Time-Dependent Hartree-Fock (TDHF) method \cite{Dir30,Sim10} to understand the mechanism responsible to the generation of angular momenta during and after the scission. Nevertheless, the present paper will not discuss the source and the amplitude of the initial fluctuations of the orientation angle of the fragments. This limits the conclusion of the present work to only qualitative results.

%This paper is organized as follows: In Sec. \ref{sec:method}, the model used to describe the generation of angular momentum is described and in Sec. \ref{sec:results} the results are discussed. Finally, the conclusion and outlooks are presented in Sec. \ref{sec:conclusion}.

 \section{Method}
 \label{sec:method}
 
 To understand the rotation during the scission,  different  TDHF trajectories are computed for a variety of scission configurations. In a three-dimensional
Cartesian  grid discretize with $N_x=N_y=$30, $N_z=130$ and a mesh spacing $dx=0.8$ fm, two fragments are placed at a distance $D$ between their center of mass on the z-axis.   The heavy (H)  and light (L) fragments are rotated to form an angle $\theta_{\rm H}$ (respectively $\theta_{\rm L}$ ) with the z-axis. The rotation is performed on the y-axis such that the principal axis of deformation of the fragments stays in the x-z plan.  When both fragments are deformed an additional rotation of an angle $\varphi$ is performed for the light fragment around the fission axis (z-axis). Combinations of   $\theta_{\rm H}$, $\theta_{\rm L}$ and $\varphi$ cover all the possible relative orientations of the fragments. The wave function of the fragments can overlap at short distances $D$, in which case the Gram-Schmidt antisymmetrization procedure is used before performing the TDHF evolution. 
The initial wave function is obtained from a static Hartree-Fock calculation; then
the initial velocity and angular momentum of the fragments start at zero.
The TDHF dynamics and static calculations are obtained using the Sly4d functional \cite{kim1997} with a simplified version of the LISE code \cite{Jin21} without pairing. The time evolution is done using the Runge-Kutta method at the order 4 with a time step $\Delta t$=0.3 fm/c.

\begin{table}[!h]
\caption{ Quadrupole and octupole deformation parameters and rigid moment of inertia of the 4 nuclei considered in this study. The rigid moment of inertia is obtained on an axis perpendicular to the main deformation axis of the nucleus. The excitation energy $E^*$ of the deformed state is also shown. Note that the deformed wave functions are in the ground state except  $^{108}$Ru which is in a local HF minimum.}
\begin{tabular}{c|c|c|c|c}
\hline
\hline
 Nuc. & $\beta_2$  &  $\beta_3$ &  $E^*$ [MeV] & $I_{\rm Rigid}$ [$\hbar^2$/MeV] \\ %
\hline
$^{132}$Sn & 0. & 0. & 0 & 50.0\\
$^{144}$Ba & 0.22 & 0.16 & 0 &  63.1  \\
\hline
$^{108}$Ru & 0.82 & 0. & 3.5 &  51.4\\
$^{96}$Sr & 0.53 & 0. & 0 & 37.1 \\
\hline
\end{tabular}
\label{tab:def}
\end{table}

  Two final state channels are studied with different deformations to disentangle the different mechanisms. The first is the  $^{132}$Sn+$^{108}$Ru channel where the heavy fragment is taken as spherical and the light one is super-deformed. The other is the  $^{144}$Ba+$^{96}$Sr for which the heavy fragment has an octupole deformation and the light one has a strong quadrupole deformation. For each system,   two cases are discussed, one at a large distance where only the Coulomb interaction makes the fission fragments interact with each other, and one at a smaller distance for which the fragments touch and form a neck that supports a nuclear interaction between the fragments.  
  
  The initial Hartree-Fock (HF) wave functions are obtained with the Sky3d code \cite{Mar14}. The Sky3d code has been slightly modified to induce an initial octupole deformation at the beginning of the HF process. By choosing carefully the initial deformation for these systems, the HF calculation converges to the final deformations shown on Tab \ref{tab:def}. The values obtained are comparable to the ones that emerge in microscopic models for the asymmetric fission of the actinide in typical static \cite{Lem19,Zha16,Mar21} and dynamical calculations \cite{Sca18,Bul19b,Qia21,Ren22}, i.e. a spherical heavy fragment or one with an octupole deformation, and a light fragment with a large quadrupole deformation.

The deformations parameters are computed as
\begin{align}
\beta_{\rm L} &= \frac{4 \pi }{3 A (r_0 A^{1/3})^{\rm L} }  Q_{l 0},\\
Q_{20}&=\sqrt{\frac{5}{16\pi}}\int d^3 {r} \,\rho(\mathbf{r}) (2z^2-x^2-y^2), \\
Q_{30}&=\sqrt{\frac{7}{16\pi}}\int d^3 {r} \,\rho(\mathbf{r}) \left( 2z^3-3z(x^2+y^2) \right),
\end{align}
 with $r_0$~=~1.2~fm.
 
  \begin{figure}[!h]
\centering
\includegraphics[width=.99\linewidth, keepaspectratio]{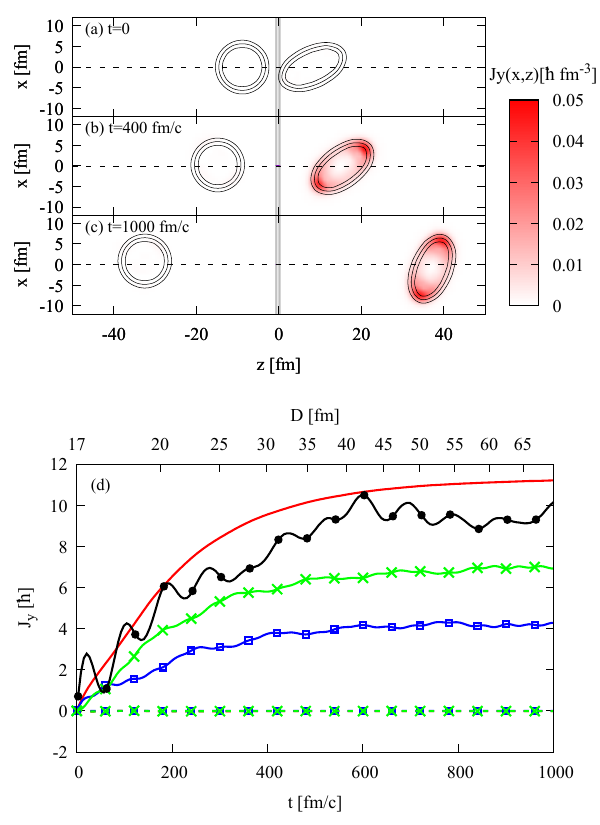}
\caption{ (Color online) Evolution of fissioning fragment $^{132}$Sn (left)+$^{108}$Ru (right)  initially at a distance $D$=17 fm. The $^{108}$Ru fragment is initially rotated by an angle $\theta$=25$^{\circ}$. 
The panels (a-c) show the contour line at densities $\rho=$ 0.03, 0.08 and 0.13 fm$^{-1}$  at the  time t= (0, 400, 1000) fm/c. The angular momentum density (the y component of Eq. \eqref{eq:j_local}) is shown by the color scale. Panel (d) : The evolution of the angular momentum as a function of time and distance $D$ is shown for the heavy fragment (dashed line) and the light fragment (continuous line).  The proton, neutron, and total angular momentum are shown with blue squares, green crosses, and red lines respectively. The black line with bullets shows the angular momentum computed with Eq. \eqref{eq:j_deriv}. In the present case, all the components of the angular momentum are close to zero for the heavy fragments.        
         } 
\label{fig:Sn_Ru_D17}
\end{figure}

  \section{Results} 
   \label{sec:results}

  \subsection{$^{240}$Pu $\rightarrow$ $^{132}$Sn+$^{108}$Ru}
   
   \subsubsection{Large distance, effect of the Coulomb repulsion}
 
Fig. \ref{fig:Sn_Ru_D17} shows the dynamics of the post-scission evolution of the fragments $^{132}$Sn+$^{108}$Ru starting at a distance $D$=17 fm and a  light fragment rotated by an angle of 25$^{\circ}$ (See Fig. \ref{fig:Sn_Ru_D17}(a)). The Coulomb repulsion produces a torque on the light fragment that causes it to rotate, as may be seen in Fig. \ref{fig:Sn_Ru_D17}(b) and (c). On the panel (b) and (c), the local angular momentum, 
   \begin{align}
  {\bf J}({\bf r}) = \hbar \sum_i  \langle \Phi_i(\bf r) |  \left( (\hat{\bf r} - {\bf r}_{cm} ) \times ( \hat{\bf p} - {\bf p}_{cm} ) +\hat {\bf s} \right )  \; {\bf \Theta_F} | \Phi_i(\bf r)  \rangle , \label{eq:j_local}
  \end{align}
 is shown for the y-direction in the y=0 plane. ${\bf r}_{\bf cm} $ and ${\bf p}_{\bf cm} $ are respectively the position and momentum of the center of mass of the fragments.  ${\bf \Theta_F}$ is the projector operator on the half-space containing the fragment $\bf F$. The division between the fragments is chosen as the plan $z$=0. Different prescriptions for the position of the plan of division would lead to small changes of the angular momentum when the fragments touch each other but give the same results when they are well separated.  On panel (d), the integral of $  {J_y}({\bf r}) $ in the whole space is shown as a function of time for protons, neutrons, and the sum of the two. The  total angular momentum can also be extracted from the angular velocity  $\dot{\theta} $  assuming that the moment of inertia $I_y$ is that of a rigid body, 
   \begin{align}
	J_y^{Rigid}(t)= I^{Rigid}_y(t) \dot{\theta}. \label{eq:j_deriv}
  \end{align}
 With $\theta(t)$ the angle between the principal axis of deformation of the fragment and the z-axis.
 As expected the spherical heavy fragment does not rotate and then has zero intrinsic angular momentum. For the light fragment, the initial tilted orientation let the Coulomb potential creates a torque in the fragment that tend to increase the $\theta_{\rm L}$ angle resulting at a large distance by a value of the intrinsic spin of 11 $\hbar$. This value is large and probably exaggerated in the present calculation due to the assumption of a large initial angle $\theta_{\rm L}$. As can be seen in Fig. \ref{fig:J_fct_theta_Sn_Ru} the final angular momentum depends strongly on the initial angle. Obviously, the torque disappears for initial values of $\theta_{\rm L}=0$ and 90$^{\circ}$.

\begin{figure}[h]
\centering
\includegraphics[width=.99\linewidth, keepaspectratio]{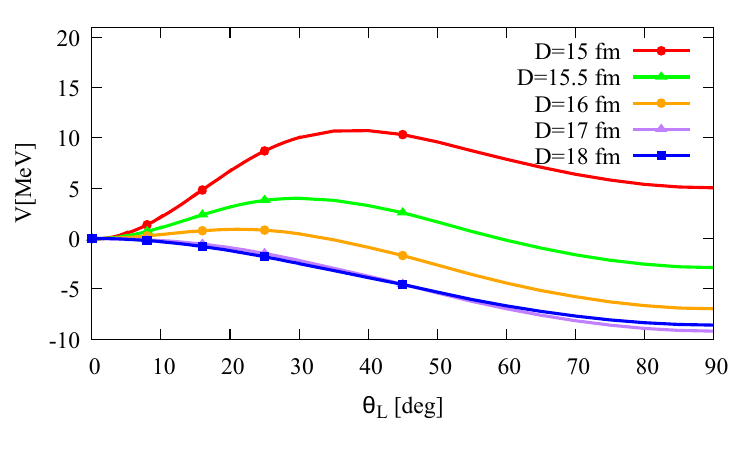}
\caption{ (Color online) 
         Frozen Hartree-Fock potential as a function of the orientation angle $\theta_{\rm L}$ of the light fragment and the distance $D$  between the center of mass of the  $^{132}$Sn and $^{108}$Ru fragments.   } 
\label{fig:pot_fct_theta}
\end{figure}

  The potential responsible for the torque is shown in Fig. \ref{fig:pot_fct_theta}. The potential is computed with the Frozen Hartree-Fock method \cite{Sim17,Uma21}. With that method, both fragments are placed in the same lattice and the energy is computed from the sum of the densities of the two fragments. Note that in the present case,  the Gram-Schmidt antisymmetrization procedure is not applied.  The antisymmetrization would modify strongly the shape of the potential at short distances making it incompatible with the corresponding dynamical results as shown in App. \ref{app:ortho}. A better method to obtain the potential would be to use the density-constrained Hartree-Fock theory \cite{Sim17} but it has been found too demanding in terms of calculations and complicated to converge in the present case. 
  
  The potential is defined to be 0 for the angle $\theta_{\rm L}=0$, 
   \begin{align}
	V(\theta_{\rm L},D) = E_{\rm FHF}(\theta_{\rm L},D) - E_{\rm FHF}(\theta_{\rm L}=0,D)
  \end{align}
 This frozen potential for $D$>17 fm shows that the $\theta_{\rm L}$= 90$^{\circ}$ configuration is more stable than the 0$^{\circ}$ one. Immediately after the scission, the difference of energy is about 10 MeV.  As the fragments move apart,  $V(\theta_{\rm L},D)$  decreases as $D^{-3}$, as expected for a  quadrupole Coulomb interaction.
 
It should be noted that the TDHF calculation preserves the expectation value of the total angular momentum.  It is initially zero since no boost is applied. The heavy fragment has no final angular momentum, and the orbital angular momentum $\vec \Lambda$ is equal to $-\vec J_{\rm L}$. This angular momentum results in a small velocity for the fragments in the x-direction, positive for the light fragment and negative for the heavy in the present case. The resulting small shift of position of the fragment in the x-direction is visible in Fig. \ref{fig:Sn_Ru_D17}(c).  In practice, due to the numerical implementation, the total angular momentum deviates less than 0.1 $\hbar$ from 0.

It is interesting to compare the TDHF results with the simpler model of ref. \cite{Ras69}.  In that model, damping of the quadrupole deformation plays an important role which reduces the final angular momentum created by the Coulomb from about 5 $\hbar$ to 2 $\hbar$. In our calculation, no damped vibration of the quadrupole deformation is found. The large quadrupole deformation persists throughout the Coulomb separation phase.

In Fig.  \ref{fig:Sn_Ru_D17}, the total $J_y$ and $J_y^{rigid}$ present two different behaviors. The first one increases continuously from 0 to a final value of 11.3 $\hbar$ while the rigid calculation oscillates and saturates at a value around 9.5 $\hbar$. The oscillations have been found to be due to the excitation of the low-energy quadrupole vibration that squeezes the nuclei in the z-axis and so changes the orientation of the principal axis. The difference between the two asymptotic values is interpreted as follows: Some parts of the angular momentum are stored in an internal motion of the nucleon inside the mean-field and do not lead to a contribution to the collective rotation of the fragment. A more pronounced effect of this non-collective rotation will be presented in the following case.

\subsubsection{Short distance, effect of the nuclear interaction }

 \begin{figure}[!h]
\centering
\includegraphics[width=.99\linewidth, keepaspectratio]{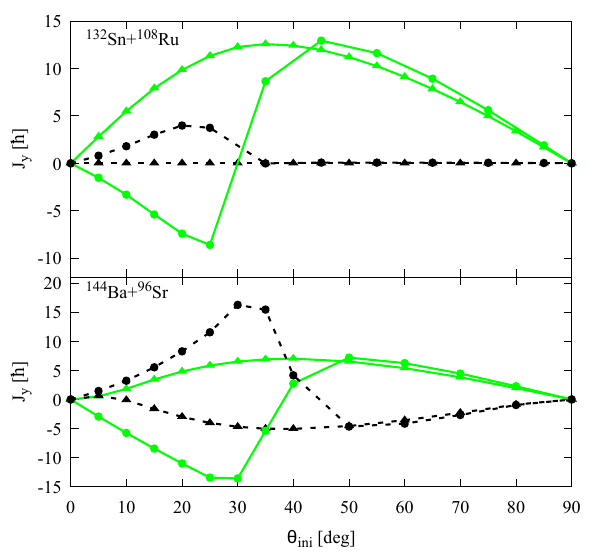}
\caption{ (Color online)  Final angular momentum in the y-axis for the heavy fragment (black dashed line) and the light fragment (green solid line) as a function of the initial angle $\theta_{\rm L}$ starting at a distance $D=17$ fm (triangles) and $D=15$ fm (bullet) in the case of the fission output $^{132}$Sn+$^{108}$Ru (top) and  $^{144}$Ba+$^{96}$Sr (bottom).
         } 
\label{fig:J_fct_theta_Sn_Ru}
\end{figure}

 \begin{figure}[!h]
\centering
\includegraphics[width=.99\linewidth, keepaspectratio]{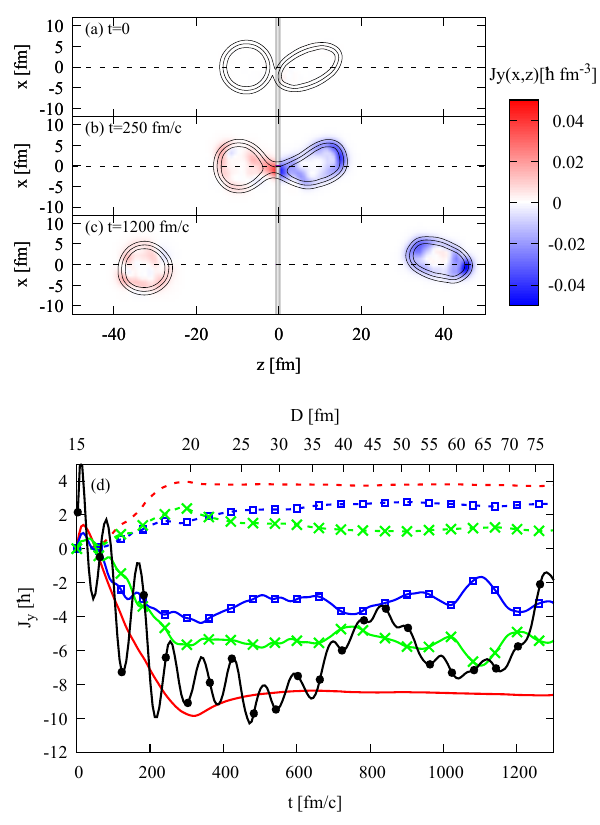}
\caption{ (Color online) Same as Fig. \ref{fig:Sn_Ru_D17} for an initial distance D=15 fm and $\theta_{\rm L}$=25$^{\circ}$. 
         } 
\label{fig:Sn_Ru_D15}
\end{figure}

The impact of the nucleus-nucleus potential is visible in Fig. \ref{fig:Sn_Ru_D15} for which the fragments are initially placed at a center of mass separation $D=15$ fm. 
 The fragments not only feel the torque due to the Coulomb repulsion but also a stronger torque in the opposite direction that is due to the attractive nucleus-nucleus interaction. The torque can be deduced from the potential of Fig. \ref{fig:pot_fct_theta} at a short distance.
The complex dynamics of the two fragments result in a bending mode\footnote{The angular momentum of each fragment are in opposite direction and perpendicular to the fission axis} with an angular momentum of 3.7 $\hbar$ for the heavy fragment and -8.6 $\hbar$ for the light. The light fragment keeps its large deformation during the process and a major part of the angular momentum is collective.   Nevertheless, the excitation of a large variety of modes due to the neck-breaking complicates the analysis of the angular momentum generated in the fragments.
Indeed, comparison with Fig. \ref{fig:Sn_Ru_D17} shows that a larger part of the angular momentum is internal to the nuclei: i) on panel (c) the angular momentum is not evenly spread on the surface of nuclei ii) the rigid angular momentum has larger oscillations iii) protons and neutrons exchange continuously about 1 $\hbar$ of angular momentum which may be due to the excitation of a scissor mode.

As in the Coulomb case, the results depend on the initial angle, as can be seen in Fig. \ref{fig:J_fct_theta_Sn_Ru}. The top panel of that figure shows that the generated angular momentum depends almost linearly on the initial angle until an angle of about 30$^{\circ}$ for which the fragment stop touching each other and so the Coulomb torque dominate leading to a situation similar to the $D$=17 fm case.

The rotation of the heavy fragment is even more peculiar. The initial fragment is spherical, but the neck and the transfer of a small number of nucleons (1.5 neutrons and 0.8 protons from the light to the heavy fragment) lead to a small deformation that breaks the spherical symmetry. Then, the rapid movement of the neck along the x-axis just before scission induces a rotation of the heavy fragment. The quasi-spherical fragment has a final angular momentum of 3.7 $\hbar$.  Surprisingly, protons contribute the most to angular momentum, although their rigid moment of inertia is lower. A slow transfer of angular momentum from neutrons to protons was also observed. No explanation for this effect has been found. The angular momentum assuming a rigid fragment is not shown in Fig. \ref{fig:Sn_Ru_D15} because shape fluctuations cause the main axis to oscillate with large amplitudes, resulting in non-physical rigid angular momentum.

   \begin{figure}[!h]
\centering
\includegraphics[width=.99\linewidth, keepaspectratio]{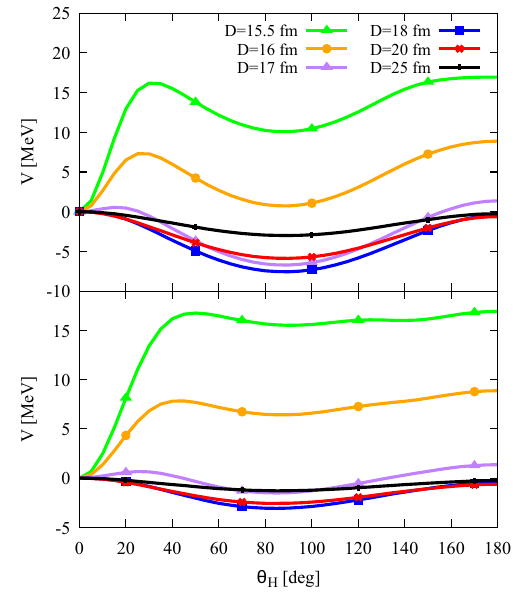}
\caption{ (Color online) 
         Frozen Hartree-Fock potential as a function of the orientation angle $\theta_{\rm H}$ of the heavy fragment and the distance $D$  between the center of mass of the  $^{144}$Ba and $^{96}$Sr fragments with $\theta_{\rm L}$=- $\theta_{\rm H}$ (top panel) and $\theta_{\rm L}$=0 (bottom panel). All of the calculations are obtained with $\varphi$=0.  } 
\label{fig:pot_fct_theta_Ba_Sr}
\end{figure} 

 \subsection{$^{240}$Pu $\rightarrow$ $^{144}$Ba+$^{96}$Sr}
 \subsubsection{Dependence of the potential with the orientation}
 
With both fragments deformed, the nucleus-nucleus potential depends on three angles, the two angles $\theta_{\rm H}$,$\theta_{\rm L}$, and the azimuthal angle $\varphi$. The potential is shown in Fig. \ref{fig:pot_fct_theta_Ba_Sr} for two cases as a function of the heavy fragment orientation angle. In the first case, both fragments are rotated with opposite angles, while in the second case only the heavy fragment is rotated. In both cases, it is interesting to note the strong effect of the octupole asymmetry of the heavy fragment. It induces a large difference of the potentials at angle $\theta_{\rm H}$=0 and $\theta_{\rm H}$=180$^{\circ}$, while in the case of pure quadrupole deformation the potential is symmetric with respect to the angle $\theta$=90$^{\circ}$.
At larger distances, the Coulomb potential is mostly unaffected by the octupole deformation.

\begin{figure}[!h]
\centering
\includegraphics[width=.99\linewidth, keepaspectratio]{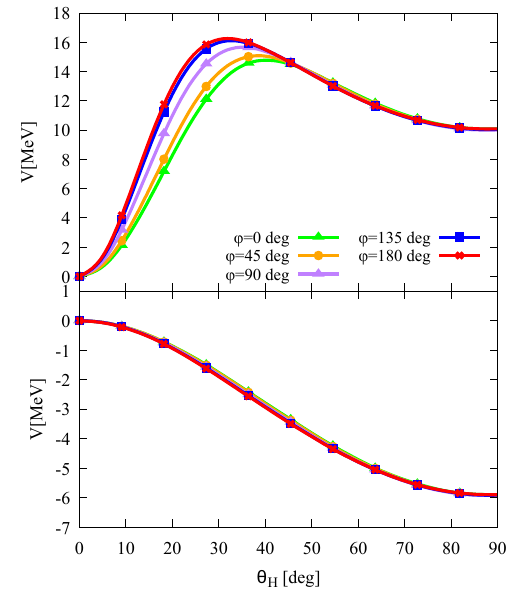}
\caption{ (Color online) 
         Frozen Hartree-Fock potential as a function of the orientation angle $\theta_{\rm H}$ of the heavy fragment taking $\theta_{\rm L}=-\theta_{\rm H}$ and several values of the azimuthal angle $\varphi$ between the  $^{144}$Ba and $^{96}$Sr fragments. The separation distance is $D$=15.5 fm (20 fm) for the upper (lower) panel respectively. } 
\label{fig:pot_fct_theta_phi_Ba_Sr}
\end{figure}

At short distances, the potential is strongly dependent on both $\theta_{\rm H}$ and  $\theta_{\rm L}$ and can not be separated into two independent contributions of each fragment. Furthermore, the azimuthal angle $\varphi$ can change the energy up to a few MeV as can be seen in Fig. \ref{fig:pot_fct_theta_phi_Ba_Sr}.  However, note that the dependence on $\varphi$ is significant only at the shorter separation and only for a limited range of $\theta_{\rm H}(=-\theta_{\rm L})$.

At large distances, where only the Coulomb interaction plays a role,  the azimuthal angle $\varphi$ does not affect much the energy of the system (see lower panel of Fig. \ref{fig:pot_fct_theta_phi_Ba_Sr}).  Similarly, the energy of the system can be well described by the independent contribution of the two fragments, i.e. the deviation,
 \begin{align}
 \left| V(\theta_{\rm H},\theta_{\rm L}) -  V(\theta_{\rm H},0) - V(0,\theta_{\rm L}) \right|, 
 \end{align}
 never exceed 0.35 MeV while  $V(\theta_{\rm H},\theta_{\rm L})$ vary up to 6 MeV for $D=$20 fm. With
  \begin{align}  
 V(\theta_{\rm H},\theta_{\rm L}) =E_{FHF}(\theta_{\rm H},\theta_{\rm L})  - E_{FHF}(0,0). 
  \end{align}
Then, the torque due to the Coulomb interaction acting on each fragment is mainly independent of the orientation of the other fragment.

\subsubsection{Large distance, effect of the Coulomb repulsion}
 
\begin{figure}[!h]
\centering
\includegraphics[width=.99\linewidth, keepaspectratio]{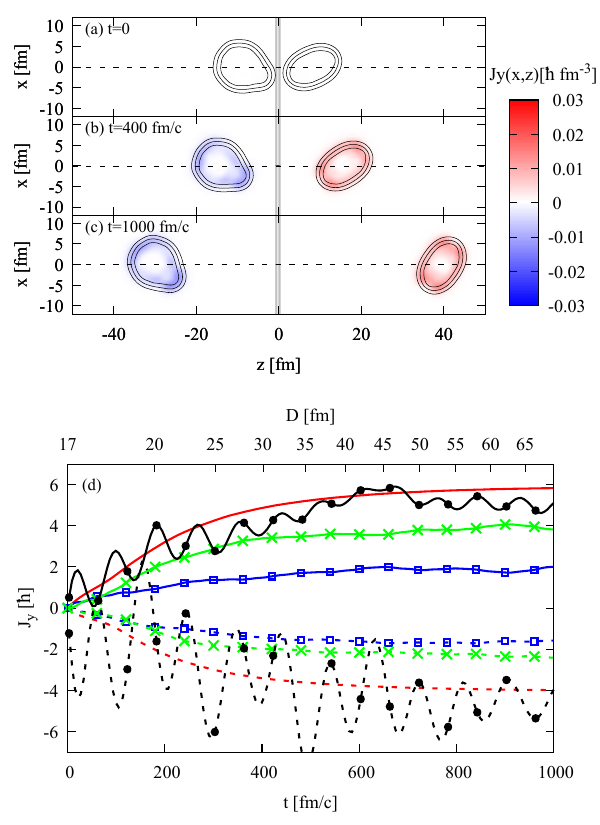}
\caption{ (Color online) Same as Fig. \ref{fig:Sn_Ru_D15} for the fission fragments $^{144}$Ba+$^{96}$Sr  with $D$=17 fm, $\theta_{\rm H}$=$-\theta_{\rm L}=25$$^{\circ}$ and $\varphi$=0.
         } 
\label{fig:Ba_Sr_D17}
\end{figure}

  The time-dependent evolution of the system  $^{144}$Ba+$^{96}$Sr at a distance $D=17$ fm with $\theta_{\rm H} = - \theta_{\rm L}$ = 25$^{\circ}$ is shown in Fig. \ref{fig:Ba_Sr_D17}.
The evolution of the light $^{96}$Sr fragment is similar to the one of the $^{108}$Ru in Fig. \ref{fig:Sn_Ru_D17}: a generation of angular momentum mainly collective.
Comparing the top and bottom panels of Fig. \ref{fig:J_fct_theta_Sn_Ru}, it can be seen that the final angular momentum is less in $^{96}$Sr than in $^{108}$Ru by about a factor of two. This may be attributed to the lower value of the quadrupole moment in $^{96}$Sr.
The heavy fragment shows also a  rotation with a final value of the angular momenta of 4 $\hbar$ due to the small quadrupole and octupole deformation.

 \begin{figure}[!h]
\centering
\includegraphics[width=.99\linewidth, keepaspectratio]{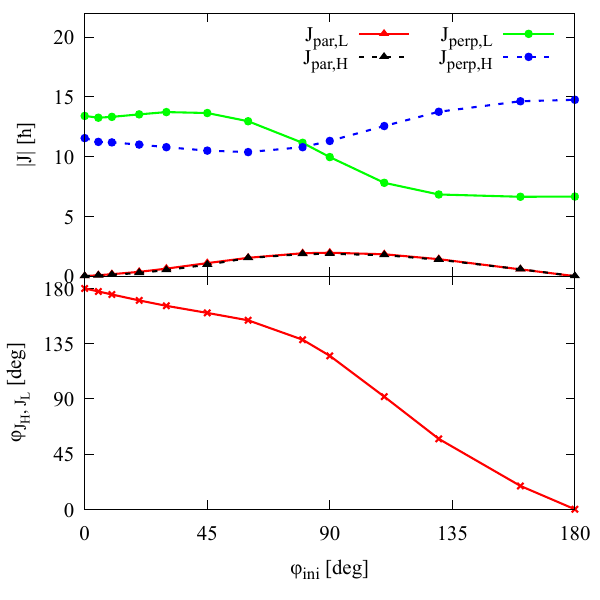}
\caption{ (Color online) Angular momentum in the direction perpendicular and parallel to the motion of the heavy and light fragment (top). The bottom panel shows the angle between the two angular momenta $\varphi_{J_{\rm H},J_{\rm L}}$ as a function of the initial azimuthal angle $\varphi$. Both panels are showing results obtained for the $^{144}$Ba+$^{96}$Sr at initial distance $D=15$ fm and $\theta_{\rm H}=-\theta_{\rm L}=25^{\circ}$.
         } 
\label{fig:J_perp_par_Ba_Sr}
\end{figure}   

 \begin{figure}[!ht]
\centering
\includegraphics[width=.99\linewidth, keepaspectratio]{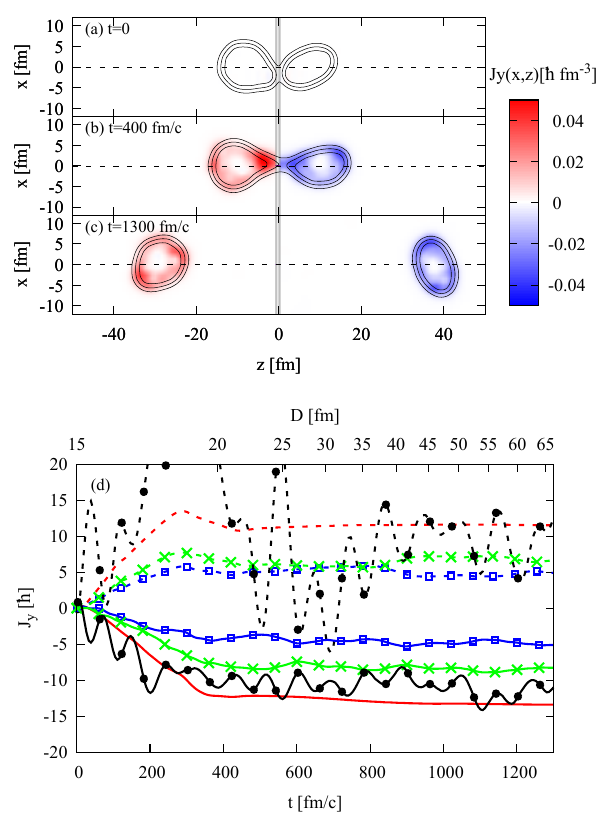}
\caption{ (Color online) Same as Fig. \ref{fig:Ba_Sr_D17} for the system $^{144}$Ba+$^{96}$Sr with $D$=15 fm, $\theta_{\rm H}$=$-\theta_{\rm L}=25^{\circ}$ and $\varphi$=0. 
         } 
\label{fig:Ba_Sr_D15}
\end{figure}

In Ref. \cite{Wil21}, it is argued that the post-scission Coulomb-induced rotation would be dependent on the product of the charge of the fragments $Z_1Z_2$ which is incompatible with the experimental data. Nevertheless, the comparison of the two fission output $^{132}$Sn+$^{108}$Ru ($Z_1Z_2$ = 2200) and  $^{144}$Ba+$^{96}$Sr ($Z_1Z_2$ = 2128) in Fig. \ref{fig:J_fct_theta_Sn_Ru} show that the main parameters that influence the generation of angular momentum by Coulomb are i) the initial angle of the fragments ii) the deformation parameters. It is difficult to estimate how the initial angle would depend on the asymmetry of the fission but it is expected that the quadrupole would have a saw-tooth shape as a function of the asymmetry which would make it compatible with the experimental data of Ref \cite{Wil21}.
 
 \subsubsection{Short distance, parallel and perpendicular component of the angular momentum }

 The dependence of the potential on the angle $\varphi$ shown in Fig. \ref{fig:pot_fct_theta_phi_Ba_Sr} is inducing a twisting\footnote{The angular momentum of each fragment are in opposite direction and parallel to the fission axis} rotational mode if the initial $\varphi$ angle is different from 0 or 180$^{\circ}$. This effect is shown in Fig. \ref{fig:J_perp_par_Ba_Sr}, where calculations are done as a function of the azimuthal angle at initial distance $D=15$ fm and $\theta_{\rm H}=-\theta_{\rm L}=25^{\circ}$. For each fragment, the angular momentum is decomposed into two components, one parallel to the motion of the fragment and one perpendicular to it. The parallel component describes the twisting component of the rotation.  A maximum value of 2.0 $\hbar$ is obtained for $\varphi$=90$^{\circ}$. This is much smaller than the perpendicular component with an average of around 10 $\hbar$. The smaller angular momentum in the twisting mode is due to: i) the gradient of the nuclear potential is smaller in the $\varphi$ direction than in $\theta$ as shown in Fig. \ref{fig:pot_fct_theta_phi_Ba_Sr} ii) the Coulomb potential very weakly contributes to the twisting mode after scission. The parallel component is purely a twisting mode (the perpendicular components of the fragment have the same norm, oriented in opposite directions) since the total angular momentum about the fission axis is conserved. 

The angle between the angular momentum $\varphi_{J_{\rm H},J_{\rm L}}$ is shown on the bottom panel of Fig. \ref{fig:J_perp_par_Ba_Sr}. It shows that the rotational mode is purely a bending mode for $\varphi$=0$^{\circ}$ to a purely wriggling\footnote{The angular momenta of the fragments are in the same directions and perpendicular to the fission axis} mode for $\varphi$=180$^{\circ}$.  Between these two values, the function of $\varphi_{J_{\rm H},J_{\rm L}}$ is not linear but tend to stay closer to the 180$^{\circ}$ value.  In a statistical approach, assuming a constant distribution of angle $\varphi$ between 0 to 180$^{\circ}$ would lead to a distribution of  $\varphi_{J_{\rm H},J_{\rm L}}$ populating more the final states with angle  $\varphi_{J_{\rm H},J_{\rm L}}$ > 90$^{\circ}$. This result is in qualitative agreement with the results found in Ref. \cite{Bul22,Bul22b}.

In Fig. \ref{fig:Ba_Sr_D15}, similarly to the $^{132}$Sn+$^{108}$Ru case, the strong nuclear interaction creates a large torque leading to a non-collective rotation of both fragments. They gain the dominant part of their angular momentum before the scission (around $t$=400 fm/c). The Coulomb potential only contributes to a small additional angular momentum for the light fragment.

\section{Conclusion}
\label{sec:conclusion}

Using the DFT-TDDFT framework without pairing, the present paper describes the mechanisms leading to a generation of angular momentum at and after the scission as a function of the orientation of the two fragments with the fission axis and their relative azimuthal angle. With the FHF method, the potential at and after scission is obtained as a function of the orientations of the fragments exploring two fission modes of $^{240}$Pu fission with different shapes of the fragments. The TDHF calculation reveals how much angular momentum is generated in the fragments as a function of the initial orientation and how is shared between collective rotation, and non-collective orbital contributions.

The main conclusions are :
\begin{itemize}
\item the Coulomb potential  at large distances creates a mainly collective rotation of the fragments while the nuclear interaction tends to generate more complex rotations which are less collective;
\item an initial spherical nucleus such as the $^{132}$Sn can rotate while remaining quasi-spherical with a noticeably  purely non-collective angular momentum;
\item the non-collective angular momentum is seen to be transferred  between the protons and the neutrons;
\item the generated angular momentum by the Coulomb force depends more on the deformation of the fragments than on the $Z_
1Z_2$ product. This result contradicts the argument of Ref. \cite{Wil21} against the post-scission generation of angular momentum;
\item when both fragments are deformed the azimuthal angle does not play a role in the Coulomb-induced rotation and has a weak effect on the nuclear interaction producing a twisting mode.
\end{itemize}

\begin{acknowledgements}
I want to thank Aurel Bulgac and George Bertsch for interesting and fruitful discussions and for carefully read the manuscript. The funding from the US DOE, Office of Science, Grant No. DE-FG02-97ER41014 is greatly appreciated. This research used
resources of the Oak Ridge Leadership Computing Facility, which is a U.S. DOE Office of Science User Facility supported under Contract No. DE-AC05-00OR22725.
\end{acknowledgements}

\appendix

\section{Gram-Schmidt orthogonalization}

\label{app:ortho}

\begin{figure}[!h] 
\centering
\includegraphics[width=.99\linewidth, keepaspectratio]{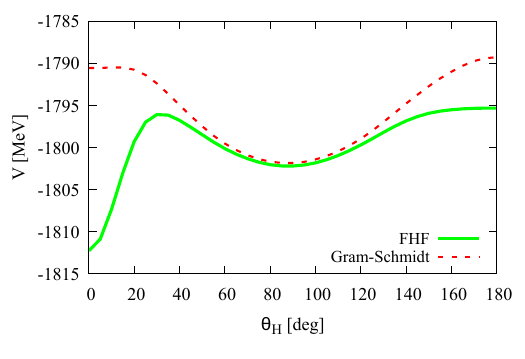}
\caption{ (Color online) Energy of the $^{144}$Ba+$^{96}$Sr at a distance $D=$15.5 fm as a function of the orientation angle of the   $^{144}$Ba with and without the orthogonalization of the wave functions.      } 
\label{fig:fig_ortho}
\end{figure}

In fig. \ref{fig:fig_ortho},  the effect of the orthogonalization on the rotational potential is shown. It can be seen that the Gram-Schmidt procedure change strongly the shape of the potential which becomes incompatible with the dynamical results presenting a strong torque for $\theta
_{\rm H}$ close to zero. This result confirms the statement of ref. \cite{Sim17} that the Gram-Schmidt should not be used with the FHF method.

 \end{document}